\documentclass[doublecol]{epl2} 
\usepackage{epsfig}
\usepackage{amssymb}
\usepackage{lscape}
\usepackage{longtable}
\usepackage{rotating}
\usepackage{amsmath}
\newcommand{\be}{\begin{equation}}
\newcommand{\ee}{\end{equation}}
\newcommand{\ba}{\begin{eqnarray}}
\newcommand{\ea}{\end{eqnarray}}

\newcommand{\bea}{\begin{eqnarray}}
\newcommand{\eea}{\end{eqnarray}}
\newcommand{\bean}{\begin{eqnarray}}
\newcommand{\eean}{\end{eqnarray}}

\title{Non-Oberbeck-Boussinesq effects in two-dimensional Rayleigh-B\'enard convection in glycerol}
\shorttitle{Non-Oberbeck-Boussinesq effects in Rayleigh-B\'enard convection}

\author{Kazuyasu Sugiyama\inst{1} 
\and Enrico Calzavarini\inst{1}
\and Siegfried Grossmann\inst{2}
\and Detlef Lohse\inst{1}}
\shortauthor{K. Sugiyama \etal}

\institute{
\inst{1} Physics of Fluids group, Department of Applied Physics, J. M. Burgers Centre for Fluid Dynamics, and Impact-, MESA-, and BMTI-Institutes, University of Twente, P. O. Box 217, 7500 AE Enschede, The Netherlands,\\
\inst{2} Fachbereich Physik der Philipps-Universitaet, Renthof 6, D-35032 Marburg, Germany
}


\abstract{
We numerically analyze Non-Oberbeck-Boussinesq (NOB) effects in two-dimensional Rayleigh-B\'enard flow in 
glycerol, which shows a dramatic change in the viscosity with temperature.
The results are presented both as functions of the Rayleigh number $Ra$ up to $10^8$ 
(for fixed temperature difference $\Delta$ between the top and bottom plates) 
and as functions of $\Delta$ (``non-Oberbeck-Boussinesqness'' or ``NOBness'') up to $50$K (for fixed $Ra$). 
For this large NOBness the center temperature $T_c$ is more than 5K larger than the arithmetic mean 
temperature $T_m$ between  top and bottom plate and only weakly depends on Ra. 
To physically account for the NOB deviations of the Nusselt numbers from its Oberbeck-Boussinesq values, 
we apply the decomposition of $Nu_{NOB}/Nu_{OB}$ into the product of 
two effects, namely first the change in the sum of the top and bottom thermal BL thicknesses, 
and second the shift of the center temperature $T_c$ as compared to $T_m$. While for water the origin 
of the $Nu$ deviation is totally dominated by the second effect 
(cf. Ahlers {\it et~al.} J. Fluid Mech. {\bf 569}, 409 (2006))
for glycerol the {\it first} effect is dominating, in spite of the large increase of $T_c$ as compared to $T_m$.
}

\begin{document}

\maketitle

\section{Introduction}\label{sec1}
In most theoretical and numerical studies on Rayleigh-B\'{e}nard (RB) convection, 
the Oberbeck-Boussinesq (OB) approximation \cite{obe79,bou03} is employed, i.e., 
the fluid material properties are assumed to be independent of 
temperature $T$ except for the density in the buoyancy term
which is taken to be linear in $T$. The problem has two control parameters, namely  
the Rayleigh number $Ra=\beta gL^3\Delta/(\kappa\nu)$
(here $\beta$ is the thermal expansion coefficient, $g$ the gravitational acceleration, 
$L$ the height, $\Delta$ the temperature difference between bottom and top plates, 
$\kappa$ the thermal diffusivity, and $\nu$ the kinematic viscosity), and 
the Prandtl number $Pr=\nu/\kappa$. For the OB case the mean temperature profile 
shows top-bottom symmetry. However, in real fluids, if $\Delta$ is  large, 
this symmetry no longer holds due to the temperature dependences of the material properties.
Thus, for given fluid, $\Delta$ appears as an additional control parameter, which characterizes 
the deviations from OB conditions, leading to so called Non-Oberbeck-Boussinesq (NOB) effects. 
The NOB signatures can be quantified by (i) a shift $T_c-T_m$ of the bulk (or center) temperature 
$T_c$ from the arithmetic mean temperature $T_m$ between the bottom and top plates) 
and (ii) by the ratio of the Nusselt numbers $Nu_{NOB}/Nu_{OB}$ in the NOB and OB cases, 
which deviates from one. Both quantities have been measured in the large $Ra$ regime 
for helium \cite{wu91a}, glycerol \cite{zha97}, ethane \cite{ahl07}, and water \cite{ahl06}
as functions of the NOB-ness $\Delta$.

As shown in Ahlers \etal \cite{ahl06} the  Nusselt number ratio $Nu_{NOB}/Nu_{OB}$ 
can be connected to $T_c$ by the identity 
\begin{equation}
\frac{Nu_{NOB}}{Nu_{OB}}= 
\frac{2\lambda_{OB}^{sl}}{\lambda_t^{sl} + \lambda_b^{sl}} \cdot  
\frac{\kappa_t \Delta_t + \kappa_b \Delta_b}{\kappa_m \Delta} 
=: F_{\lambda} \cdot F_{\Delta}.
\label{nu_ratio}
\end{equation}
Here the labels on material properties indicate the temperature at which they are taken, e.g. 
$\kappa_t = \kappa (T_t)$ etc. $\Delta_t=T_c-T_t$ and $\Delta_b=T_b-T_c$ 
denote the temperature drops over the top and bottom thermal boundary layers, 
and $\lambda_t^{sl}$ and $\lambda_b^{sl}$ indicate their thicknesses, based on the temperature
slopes at the top and bottom plates, respectively. $\lambda_{OB}^{sl}$ is the thermal BL thickness 
in the OB case, both at top and at bottom. The factor $F_\Delta$ can be calculated from the 
temperature dependences of the material properties immediately, once $T_c$ is known. 
Remarkably, Ahlers \etal \cite{ahl06} experimentally found that for water 
\begin{equation} \label{co-in}
F_\lambda \approx 1 . 
\end{equation} 
This has been confirmed by numerical simulations of 2D NOB Rayleigh-B\'enard convection in ref. \cite{sug07}.
If the relation eq.(\ref{co-in}) holds, the Nusselt number ratio $\frac{Nu_{NOB}}{Nu_{OB}}$ already 
follows from the center temperature $T_c$, which for water can be calculated within a generalized boundary 
layer theory introduced in ref. \cite{ahl06}. 

The objective of this letter is to answer the apparently important question whether the relation 
$F_\lambda\approx 1$, meaning that the sum of the boundary layer thicknesses stays the same as in the OB case also 
under NOB conditions, $\lambda_b^{sl} + \lambda_t^{sl} \approx 2 \lambda_{OB}^{sl}$, is more
generally valid, i.e., if it holds for other liquids too. We therefore have performed (two-dimensional) 
NOB simulations with glycerol as the working fluid. For glycerol the kinematic viscosity dramatically 
depends on temperature, i.e., one should expect large changes of the boundary layer thicknesses 
at top and bottom. For instance, $\nu$ decreases from $1759~{\rm mm}^2/{\rm s}$ 
to $52.5~{\rm mm}^2/{\rm s}$ if the temperature increases from $15^{\rm o}$C to $65^{\rm o}$C. 
Another advantage of considering glycerol is the existence of experimental data for the center 
temperature (see ref.\cite{zha97}) for comparison (but not for the Nusselt number modification). 
Our main result will be that the sum of the boundary layer widths is indeed changed under NOB 
conditions, i.e., relation (\ref{co-in}) does {\it not} hold for glycerol. Its validity for water 
thus turns out to be coincidental, due to the specific temperature dependences of its material 
parameters for the chosen temperatures in the experiments of ref.\ \cite{ahl06}. 

Note that the fluid flow in glycerol is very different 
from that in water at the same $Ra$.
Due to glycerol's huge 
Prandtl number of about $Pr \approx 2500$, 
 the transition range between the onset of 
convection at $Ra_c$ and the loss of spatial coherence in the flow is much more extended than 
for water or air, whose Prandtl numbers are of order one. While in air and in water this 
transitions range extends to about $Ra \approx 5 \cdot 10^7$ to $10^8$, only beyond which there is 
turbulent convection in the bulk, this range extends to much larger $Ra$ in glycerol, namely to Rayleigh numbers 
of order $Ra \approx 10^{12}$. Since the numerical calculations cover the range up to
$Ra \approx 10^8$ only, all results refer to a fluid flow still having coherent structures.  

\begin{figure}
\begin{center}
\epsfig{file=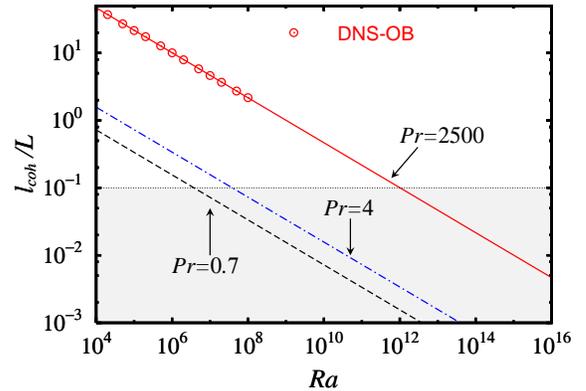,width=5.5cm,angle=-90}
\end{center}
\caption{(color online)
The coherence length $\ell_{coh}$ in multiples of the cell size $L$ versus $Ra$ number for three fluids. 
For air and water it is of order $0.1$ near $Ra \approx 10^7$ to $Ra \approx 10^8$. For glycerol this 
is reached much later; turbulent heat convection is expected beyond $Ra \approx 10^{12}$ only.
Gray shaded region indicates the developed turbulent regime. Lines are derived from the unified theory of ref.\ \cite{gro00},
symbols correspond to the Rayleigh numbers of 
the present OB numerical simulations.
} 
\label{coherence-vs-ra}
\end{figure}

To quantify these statements we use an averaged
 Kolmogorov length $\eta_K$ 
as a measure for the 
scale of coherent structures in the flow, more precisely
\begin{equation} \label{kolmogorov} 
\ell_{coh} = 10 ~\eta_K  = 10 ~(\nu^3 / \varepsilon_u  )^{1/4}.
\end{equation}
Here $\varepsilon_u$ is the volume average of the energy
dissipation rate of the flow for which
the well-known exact relation 
$\varepsilon_u  = \nu^3 L^{-4} Pr^{-2} Ra (Nu -1)$ holds.
With this we obtain
\begin{equation} \label{coherence} 
\ell_{coh} / L = 10 ~Pr^{1/2} \left( Ra (Nu - 1) \right)^{-1/4} 
\end{equation}
as an estimate for a  volume-averaged relative coherence length.
Taking $Nu(Ra,Pr)$ from the unified theory 
of refs.\ \cite{gro00}, 
one thus obtains an estimate of the coherence 
length as a function of $Ra$ and $Pr$ from eq.\ 
(\ref{coherence}), see figure 
\ref{coherence-vs-ra}.
The main features of the coherence length are (i) its pronounced 
explicit dependence  on $Pr$ (the implicit dependence via $Nu$ is only weak).
It is by about a factor $\sqrt{2500} = 50$ larger for glycerol than for gases or water. (ii) Its $Ra$-dependence is 
approximately $\ell_{coh} \propto Ra^{-0.3}$.

\section{General description of numerical simulation} \label{sec2}
We numerically solve the incompressible $(\partial_i u_i=0)$ Navier-Stokes equations
\begin{equation}
\begin{split}
\rho_m (\partial_t u_i + u_j \partial_j u_i)
=&
-\partial_i p  + \partial_j (\eta ( \partial_j u_i +\partial_i u_j ))
\\&
+g\left(\rho_m -\rho \right)  \delta_{i3},
\end{split}
\label{ns}
\end{equation}
and the heat-transfer equation
\begin{equation}
\rho_m c_{p,m} 
(\partial_t T + u_j \partial_j T ) = 
\partial_j ( \Lambda \partial_j  T ).
\end{equation}
The temperature dependence of the dynamic viscosity $\eta(T)$, the heat conductivity $\Lambda(T)$, and the
density $\rho$ are experimentally known for glycerol. They are given in the appendix of reference \cite{ahl06}.
As justified in that reference, we can assume the isobaric specific heat capacity $c_p$ and the 
density $\rho$ in the time derivatives of the material parameters to be constant at their values 
$\rho_m$ and $c_{p,m}$ at the arithmetic mean temperature $T_m$. We vary the Rayleigh number $Ra$ up 
to  $10^8$ and the level of the NOBness $\Delta$ up to $50$K. 

The container is two-dimensional (2D, no $y$-dependence), has height $L$, and aspect ratio $1$. The flow 
is wall-bounded, i.e., we use no-slip boundary conditions at all solid boundaries: $u_i=0$ at the top 
($z=L$) and bottom ($z=0$) plates as well as on the side walls $x=0$ and $x=L$. For the temperature at the 
side walls heat-insulating conditions are employed and $T_b-T_t = \Delta$ is the temperature drop across 
the whole cell. The Rayleigh number is defined with the material parameters taken at the mean temperature 
$T_m$, i.e., $Ra =\frac{\beta_m g L^3 \Delta}{\nu_m \kappa_m}$. The arithmetic mean temperature is fixed 
at $T_m=40^{\rm o}{\rm C}$. We vary the Rayleigh number by varying the height $L$ of the box, while the 
NOBness is changed by varying the temperature drop $\Delta$. Note that in the buoyancy term in eq.(\ref{ns}) 
the full temperature dependence 
of the density is taken into account, rather than employing the linear approximation 
$\rho(T)-\rho_m = \rho_m \beta (T-T_m)$ only. (Nevertheless, the Rayleigh number is defined as usual with the 
linear expansion coefficient of the density with respect to temperature, taken at $T_m$, namely 
$\beta_m = - {1\over \rho_m} {d \rho \over dT}|_{T_m}$.) The Prandtl number is defined as 
$Pr = \nu_m /\kappa_m$; for glycerol at the chosen temperature $T_m$ its value is $Pr=2495$. 
The basic equations are directly solved on the two-dimensional domain by means of the fourth-order 
finite difference method. For a detailed description of the simulation method as well as its
validations, see ref.\cite{sug07}.

One may worry if two-dimensional simulations are sufficient to reflect the dynamics of the three-dimensional 
RB convection. For convection under OB conditions this point has been analyzed in detail in ref.\ \cite{sch04} 
and earlier in refs.\ \cite{delu90,wer91,wer93,bur03}. The conclusion 
 is that for $Pr\ge 1$ 
various properties observed in numerical 3D convection and in experiment
 are well reflected in 
2D simulations. This in particular holds for the BL profiles and for the Nusselt number. 
Since the focus of this paper is on the difference between OB and NOB convection, the restriction to 2D 
simulations seems to be even less severe, as NOB deviations are expected to be similar in both 2D and 3D 
simulations and remaining differences to cancel out in quantities such as $T_c - T_m$ or $Nu_{NOB}/Nu_{OB}$.

\begin{figure}
\begin{center}
\epsfig{file=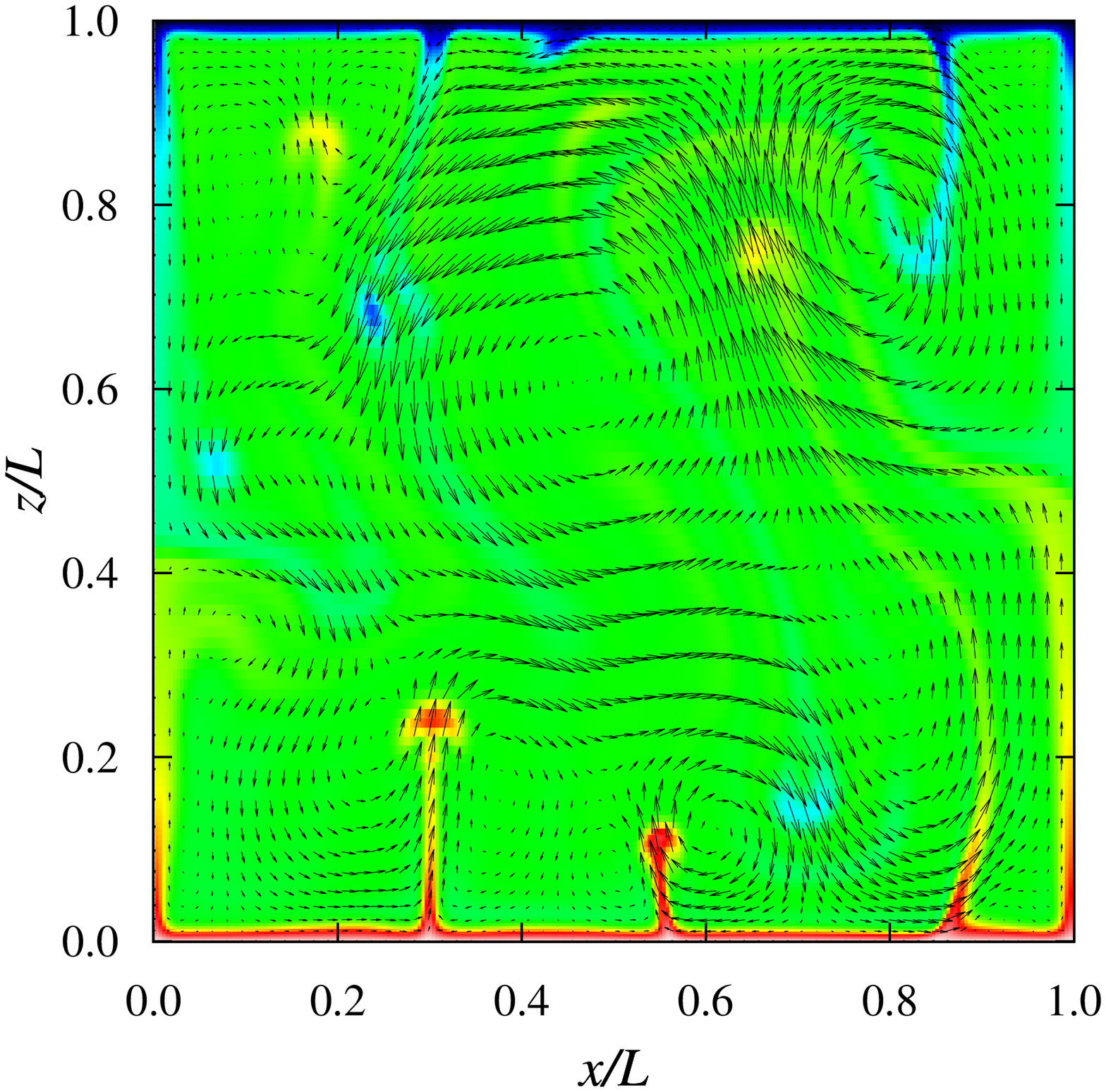,width=6.5cm,angle=-90}
\epsfig{file=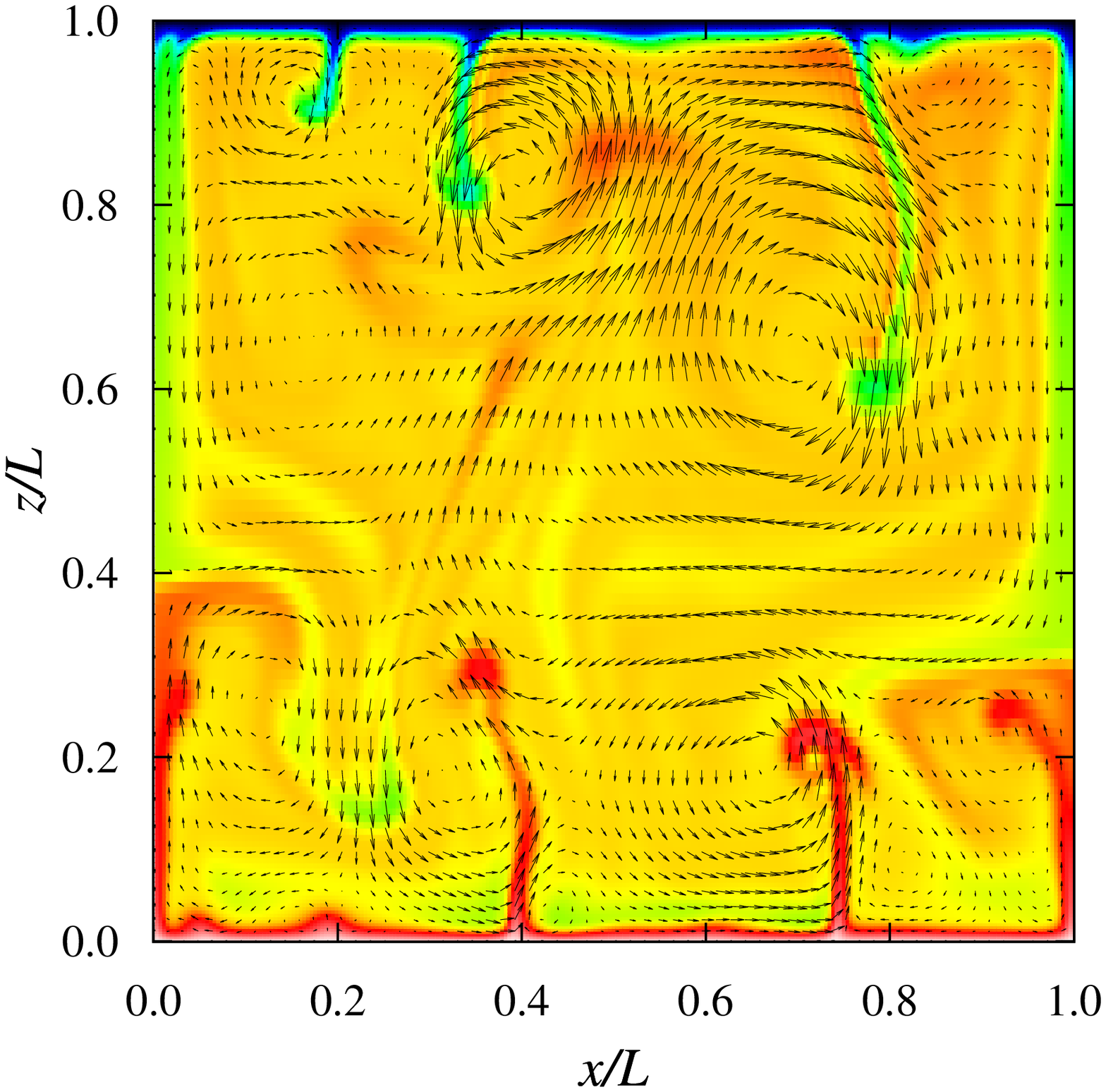,width=6.5cm,angle=-90}
\end{center}
\begin{flushright}
\epsfig{file=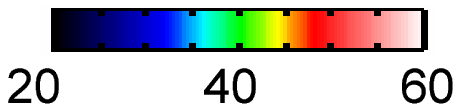,width=2.5cm,angle=0}$\ ^{\rm o}{\rm C}$
\end{flushright}
\caption{(color online)
Snapshots of the velocity and temperature fields for $Ra=10^8$ at $T_m=40^{\rm o}{\rm C}$. 
The upper panel corresponds to the OB case ($T$-independent material parameters), the lower one 
to the NOB case, both with $\Delta = 40{\rm K}$. The temperature color scheme is the same in both cases.
In the NOB case a strong temperature enhancement of the center is clearly visible. 
} 
\label{snap}
\end{figure}

\section{Results and discussions} \label{sec4}
\subsection{Large scale flow dynamics and temperature snapshots}
In the steady flow regime ($Ra<1.5 \cdot 10^5$)
a single large-scale circulation role develops, which however disappears 
in the unsteady flow regime ($Ra>1.5\cdot 10^5$) and  does not reappear up to 
the largest accessible value $Ra=10^8$ of the present study.
Even if we start the simulation with an artificial single roll, 
the large-scale circulation disappears in the course of time
and then isolated plumes (as shown in figure \ref{snap}) dominate the flow.
This feature holds for both cases, OB and NOB, and is qualitatively different from the observations in
2D (OB and NOB) simulations in water (see ref.\cite{sug07}). We attribute this to the much larger spatial
correlations in glycerol as addressed above. Note that in experiment (ref. \cite{zha97}) for 
 larger 
$Ra=2.3\cdot 10^8$ a large-scale 3D circulation role has been observed for glycerol. 
The different behavior
 between the present DNS and the experiment could either be 
due to the smaller $Ra$ or to the two-dimensionality in the simulation.

Typical temperature snapshots are shown in Figure \ref{snap}. As observed in experiments, refs.\cite{ahl06,zha97}, 
the NOB convection is characterized by an enhancement of the bulk temperature $T_c$, and a top-bottom asymmetry
of the thermal BL thicknesses. Due to the large variation of the glycerol viscosity 
(the viscosity ratio reaches as much as $\nu_t/\nu_b\approx 16$ at $\Delta=40$K),
the more viscous cold plumes from the top BL are much less mobile than the warmer plumes from the bottom 
BL. This results in a significant increase of $T_c$ as compared to the water case.

\begin{figure}
\begin{center}
\vspace*{0.3cm}
\epsfig{file=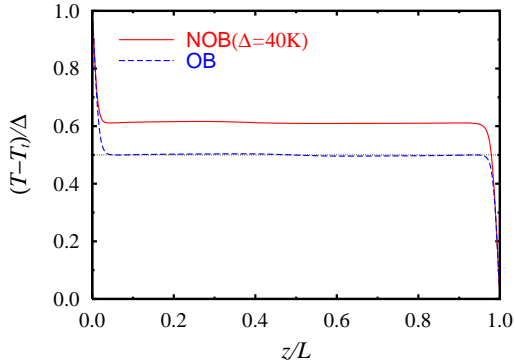,width=5cm,angle=-90}
\vspace*{0.3cm}
\end{center}
\caption{(color online)
Mean temperature profiles for glycerol at $Ra=10^8$ in the OB case (dashed) and
in the NOB case with $\Delta = 40K$ (solid). (In both cases $T_m=40^o$C, same $Ra$, same $\Delta$, but 
$T$-independent (OB) or $T$-dependent (NOB) material parameters, respectively.) In the NOB case the strong 
temperature enhancement of the center temperature $T_c$ by about 5$~K$ becomes visible 
(relative shift $\approx 0.12$). 
} 
\label{glyc_prof}
\end{figure}

\subsection{Mean temperature profiles and center temperature}
To quantify the enhancement of the bulk temperature $T_c$, the temperature profiles for $Ra=10^8$ are 
shown in Fig. \ref{glyc_prof}. Again, a strong asymmetry between top and bottom is observed:
Due to the more mobile bottom plumes the center temperature $T_c$ is significantly larger than $T_m$.

\begin{figure}
\begin{center}
\epsfig{file=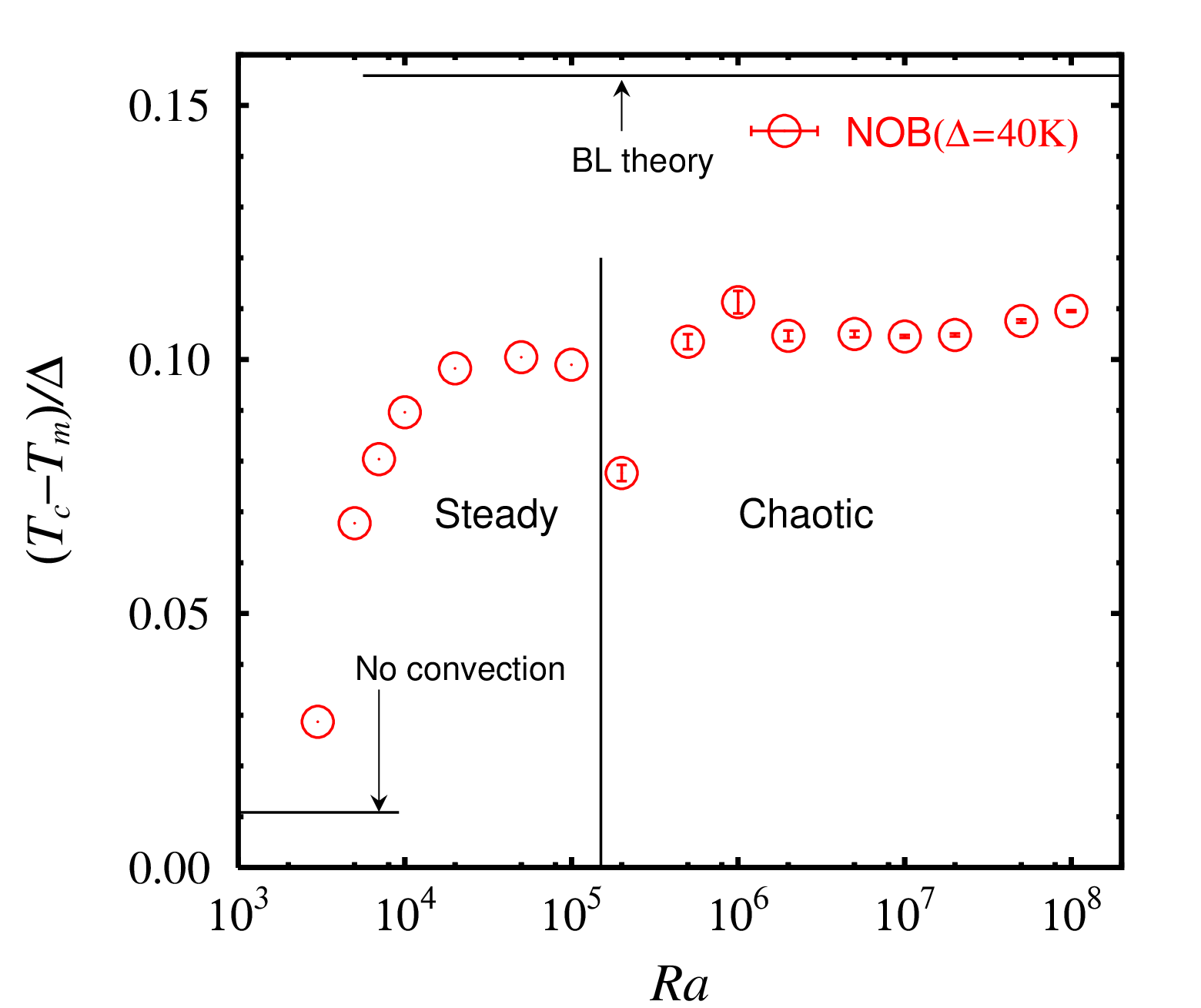,width=5.5cm,angle=-90}
\epsfig{file=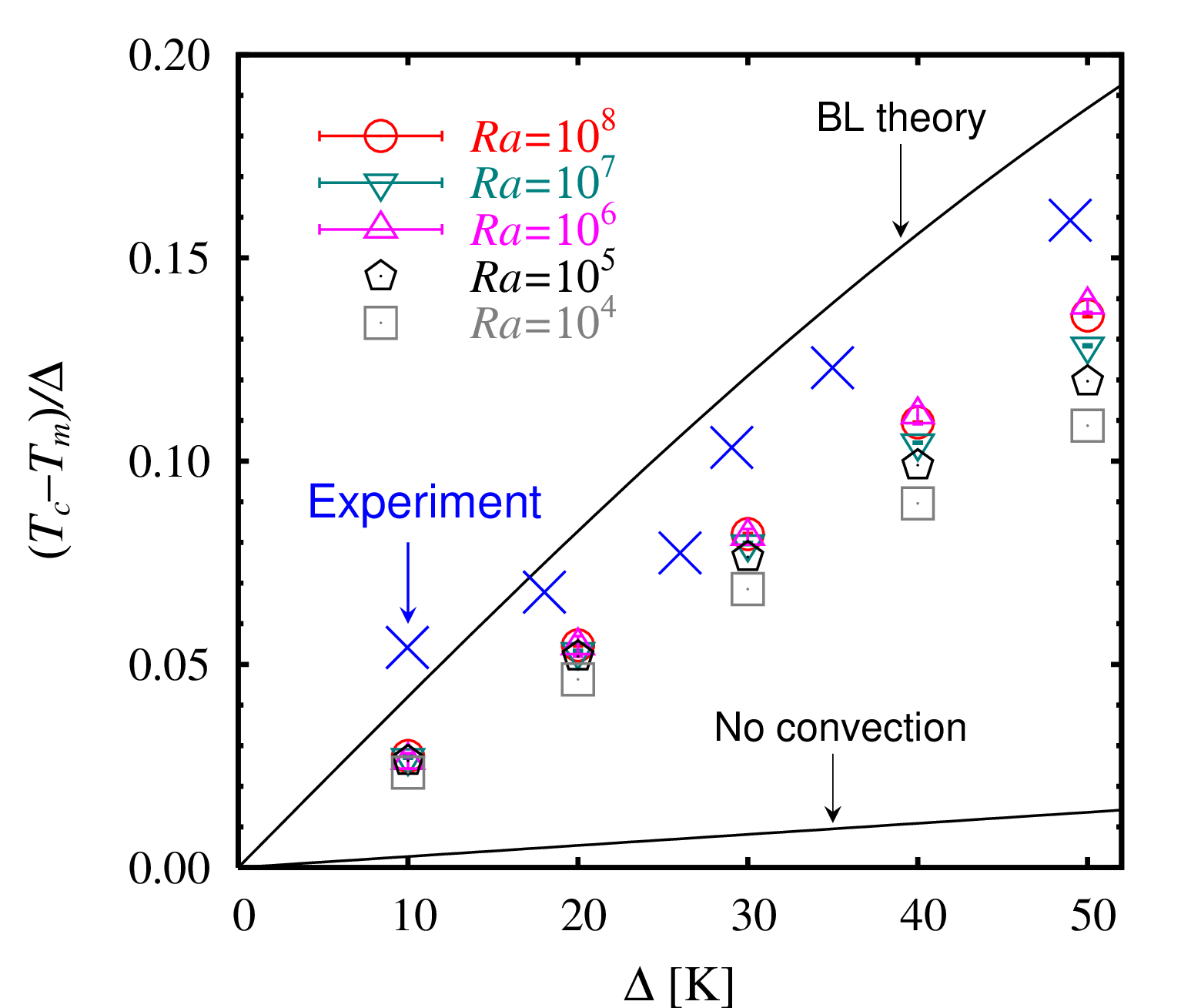,width=5.5cm,angle=-90}
\end{center}
\caption{(color online) 
Relative deviation $(T_c-T_m)/\Delta$ of the center temperature $T_c$ from the arithmetic mean temperature $T_m$
for glycerol versus $Ra$ at fixed $\Delta = 40K$ (upper) and versus the NOBness $\Delta$ at fixed $T_m=40^o$C 
and various values of $Ra$ (lower). The experimental data points
(denoted by $\times$) are measured at $Ra=2.3 \cdot 10^8$ and 
were taken from ref. \cite{zha97}. For comparison the $T_c$ shift obtained from BL theory (upper solid lines) 
and for the case of no convection (lower solid lines) are also plotted.  
} 
\label{glyc-tc-vs-del}
\end{figure}

To demonstrate this the center temperature shift $T_c-T_m$ (normalized by $\Delta$)
as function of the Rayleigh number $Ra$ and of the NOBness $\Delta$ is shown in Figure \ref{glyc-tc-vs-del}.
Except for small Rayleigh numbers just above onset of convection and in a region around 
$Ra\approx 2\cdot 10^5$ just above the onset of unsteady motion, the bulk temperature shift 
$(T_c-T_m)/\Delta$ is rather independent of $Ra$. The tiny increase between $Ra=10^7$ and $10^8$ however 
is beyond the statistical error-bars. For comparison, the prediction of the NOB BL theory given in 
ref. \cite{ahl06} and the shift for the non-convective state (i.e., purely conductive heat transport, 
driven by the temperature gradient only) are shown. Though the NOB BL theory from ref.\cite{ahl06} is 
not applicable here due to the lack of a large scale wind, it gives the correct qualitative trend for the shift 
$(T_c-T_m)/\Delta$. We also included experimental data measured at $Ra= 2.3 \cdot 10^8$ (taken from ref.\
\cite{zha97}) in an aspect ratio 1 cylindrical container. Though for that case a large scale convection 
role has been observed, the agreement with the 2D numerical simulations is reasonable.

\begin{figure}
\begin{center}
\epsfig{file=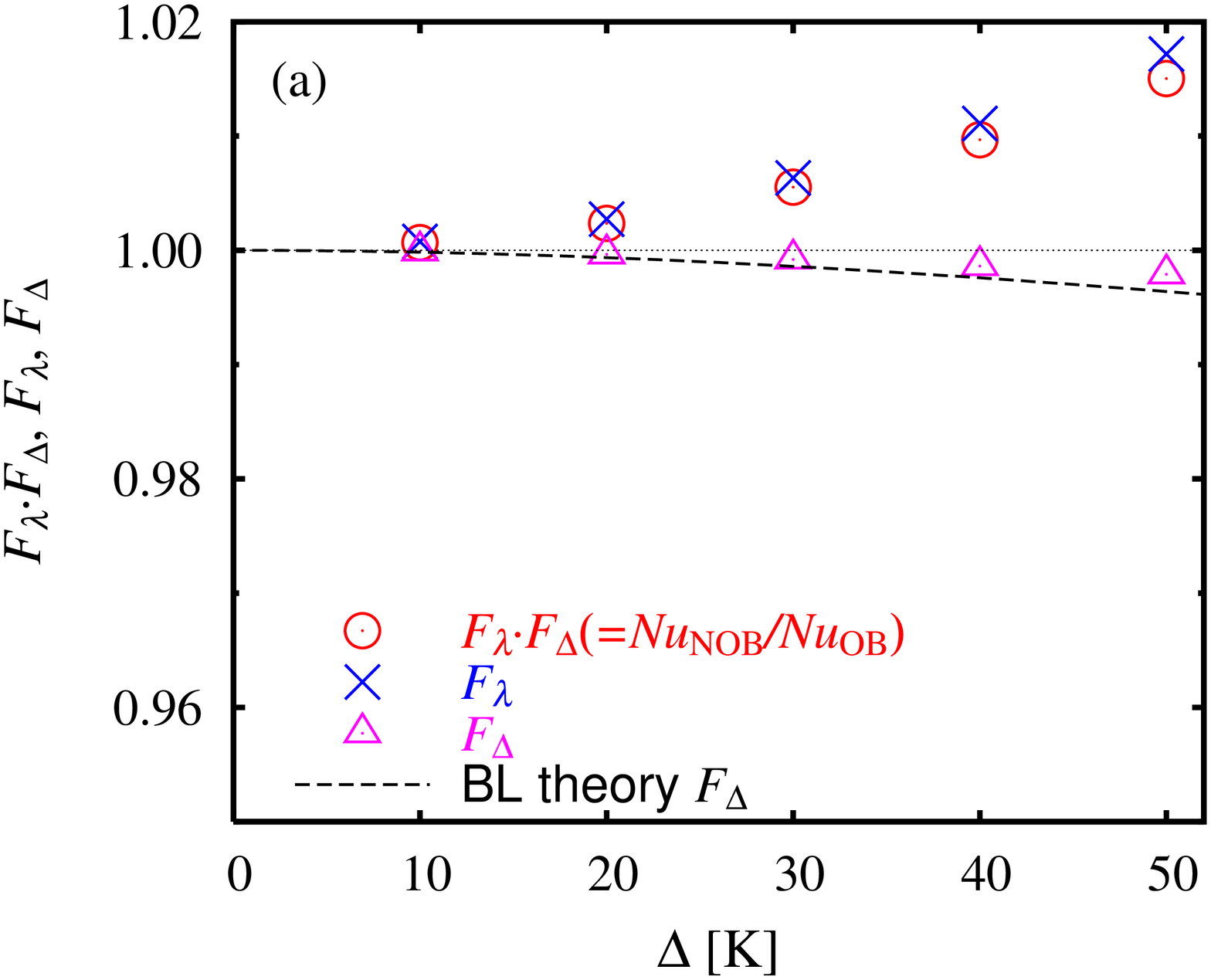,width=5cm,angle=-90}
\epsfig{file=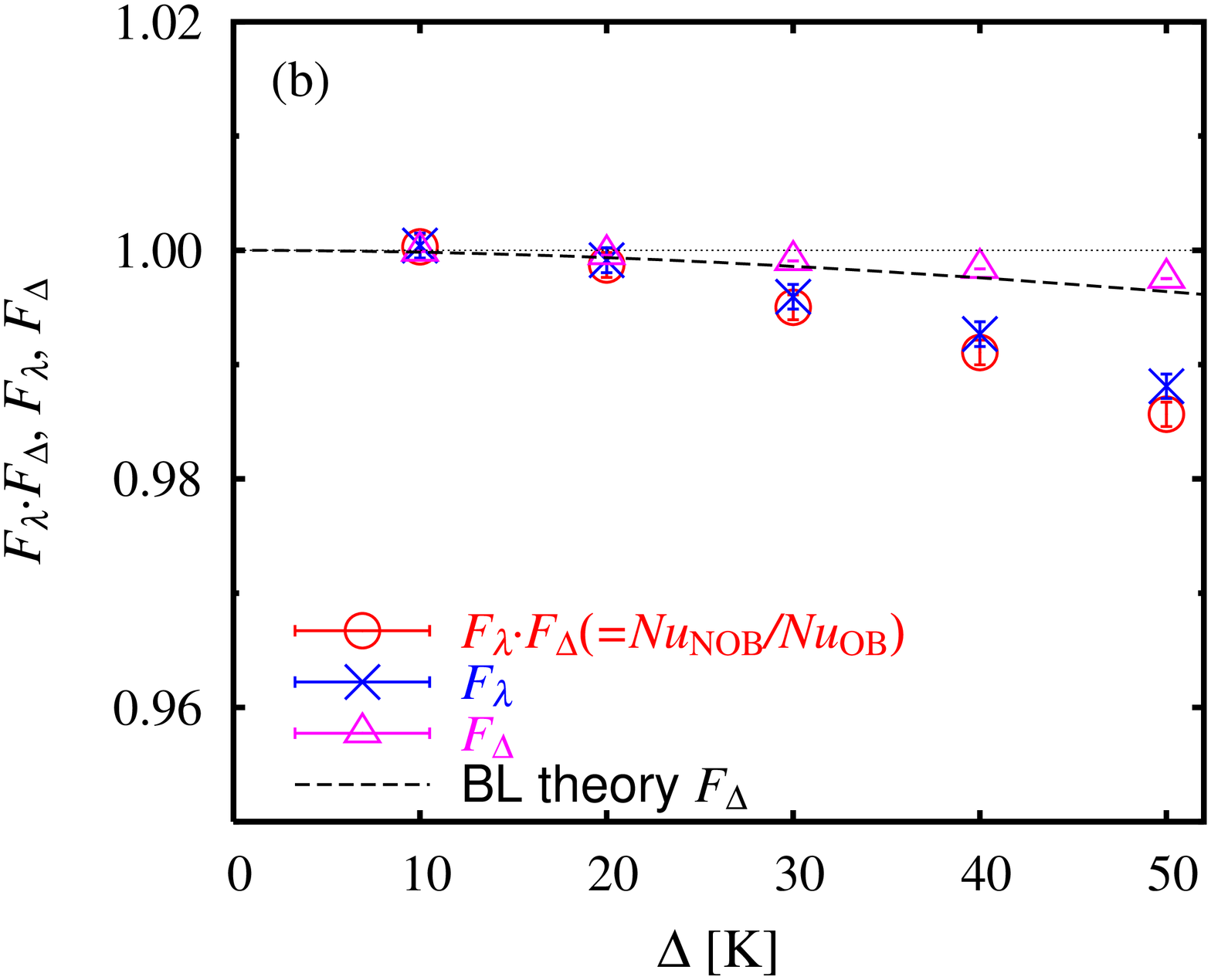,width=5cm,angle=-90}
\epsfig{file=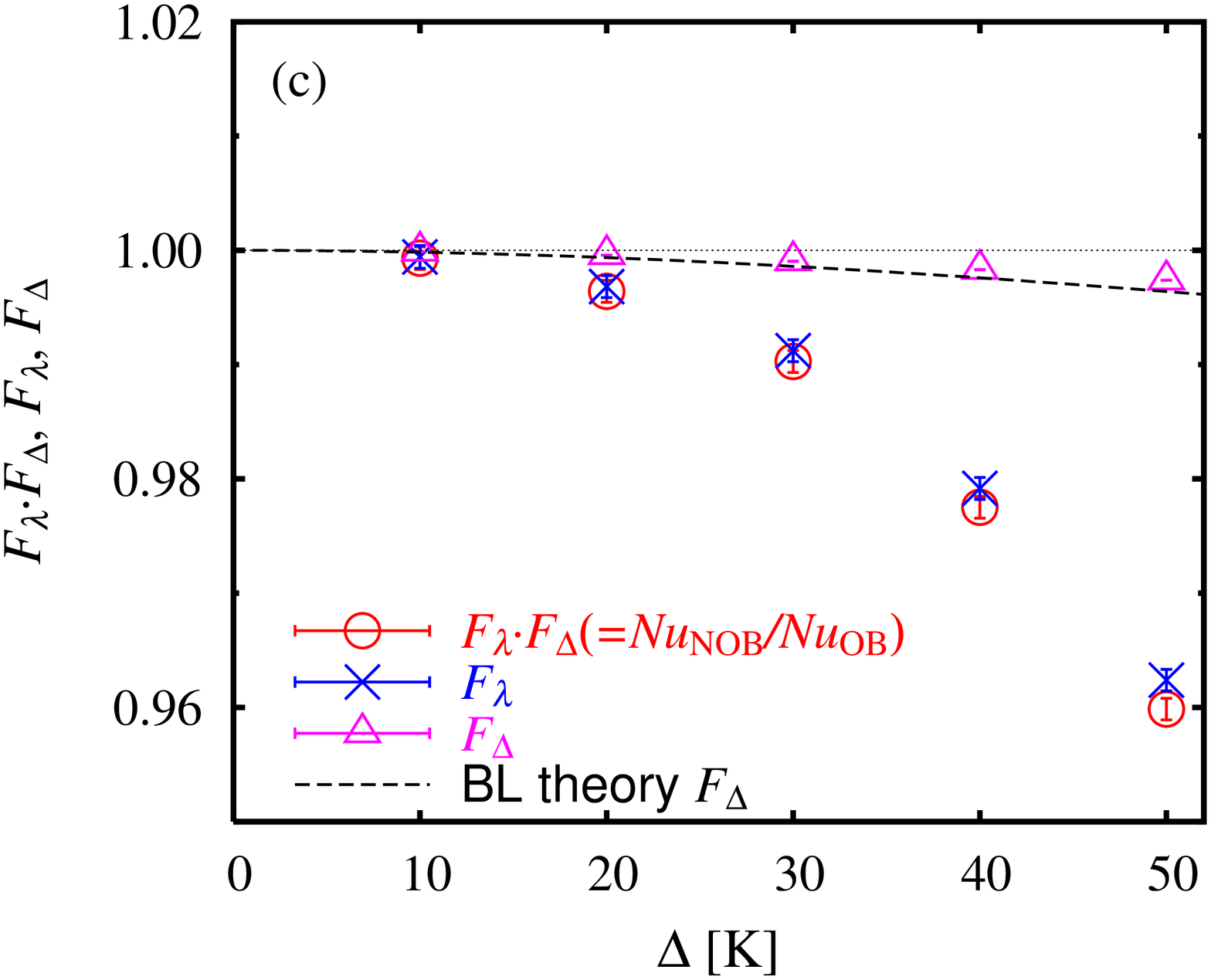,width=5cm,angle=-90}
\end{center}
\caption{(color online) 
Nusselt number ratio $Nu_{NOB}/Nu_{OB} = F_{\lambda} \cdot F_{\Delta}$ together with its contributing 
factors $F_{\lambda}$ and $F_{\Delta}$ versus \ $\Delta$ for fixed Rayleigh numbers. (a) $Ra=10^4$, 
(b) $Ra=10^7$, and (c) $Ra=10^8$. As always, the working liquid is glycerol at $T_m = 40^o$C. The dashed 
lines correspond to $F_{\Delta}$ resulting from the NOB BL theory of ref.\cite{ahl06}.
}
\label{glyc-f-vs-del}
\end{figure}

\begin{figure}
\begin{center}
\epsfig{file=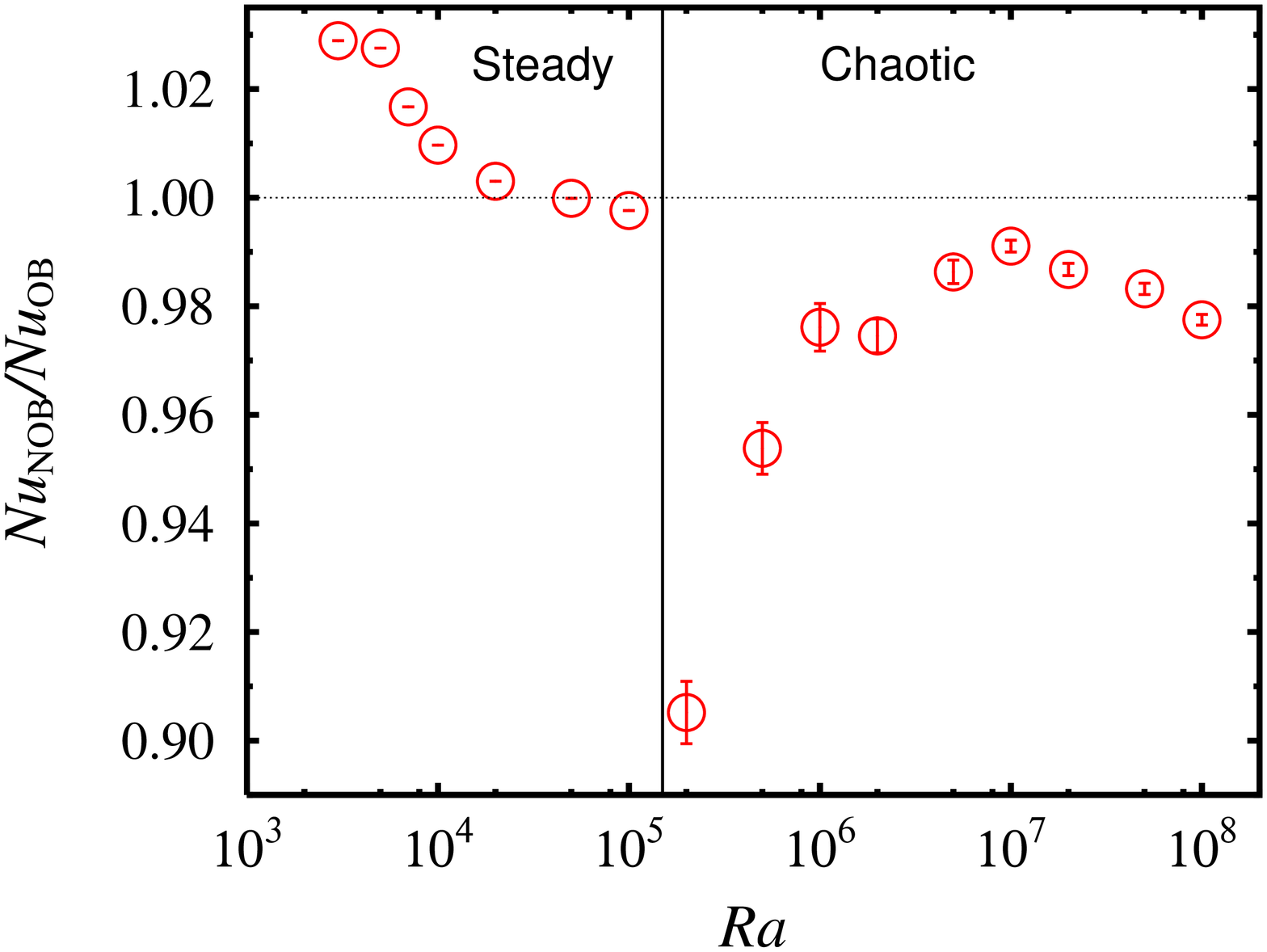,width=5.0cm,angle=-90}
\epsfig{file=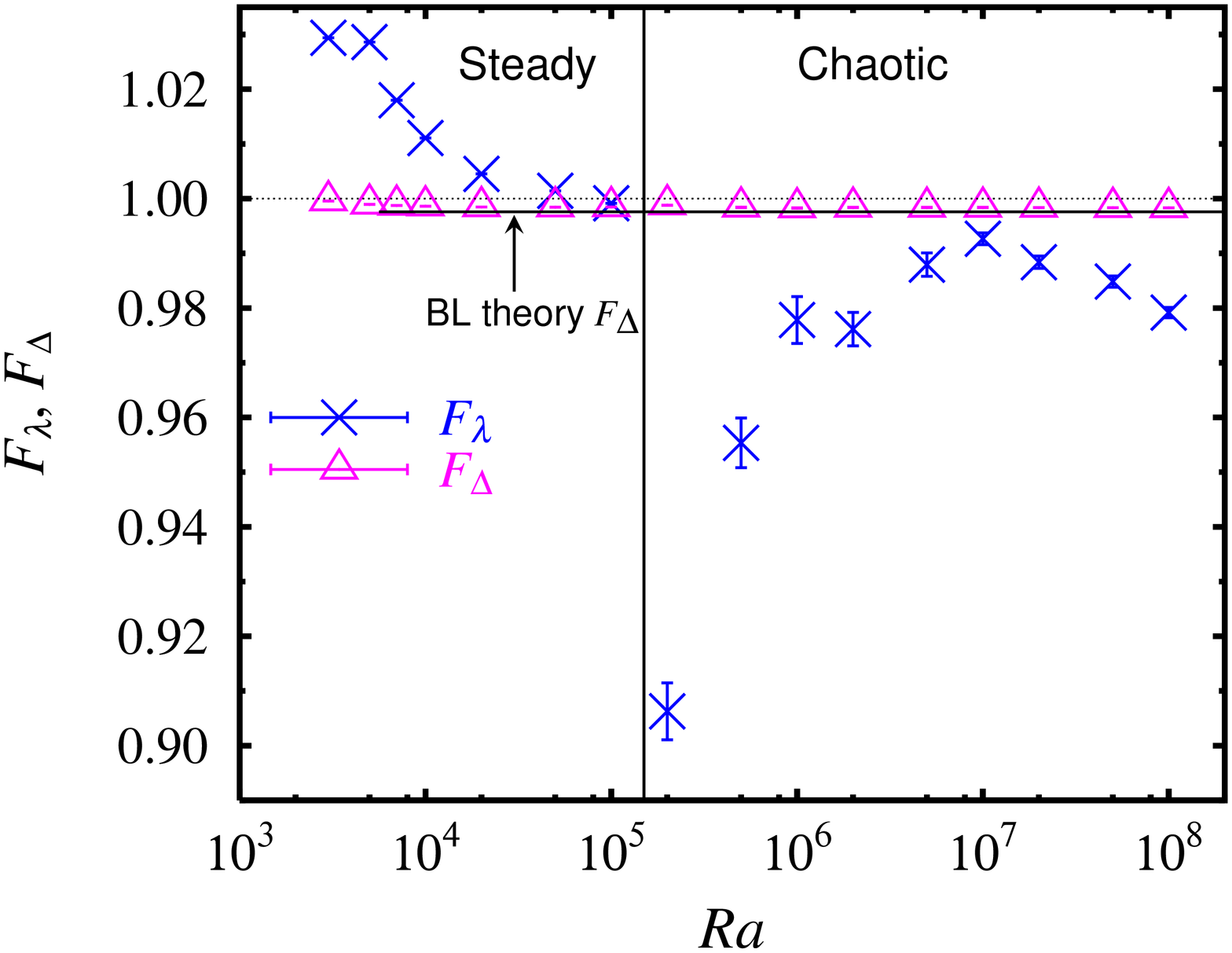,width=5.0cm,angle=-90}
\end{center}
\caption{(color online) 
The Nusselt number ratio $Nu_{NOB}/ Nu_{OB} = F_\lambda \cdot F_\Delta$ (upper) and the constituting factors
$F_{\lambda}$ and $F_{\Delta}$ individually (lower) versus $Ra$ for fixed NOBness $\Delta = 40K$ (glycerol 
at $T_m = 40^o$C). 
Note the dramatic NOB effect at $Ra\approx 2\cdot 10^5$; this happens still in the pattern forming 
range, far below the turbulent high $Ra$ region. We are not aware of its experimental verification.
} 
\label{glyc-nu-vs-del}
\end{figure}

\subsection{Nusselt number} 
The key question on NOB effects is: How do they affect the heat flux, i.e., the Nusselt number? 
For water we could address this question within an extended BL theory, cf. ref.\cite{ahl06}, but only thanks 
to the exact relation eq.(\ref{nu_ratio}) and the {\it experimental} input $F_\lambda \approx 1$, see relation 
(\ref{co-in}), because then only $F_\Delta$ is needed to calculate the NOB deviations in the Nusselt number ratio,
and $F_\Delta$
 is accessible within the extended BL theory, since 
it follows directly from $T_c$.
But here, with glycerol as working fluid, we find that $F_\lambda \approx 1$ does {\it not} hold, as 
demonstrated in Fig.\ \ref{glyc-f-vs-del}. In contrast to water, for glycerol 
the main $\Delta$-dependence of  $Nu_{NOB}/Nu_{OB}= F_\lambda \cdot F_\Delta$ is due to the $\Delta$-dependence 
of $F_\lambda$ while the factor $F_\Delta$ is basically 1 for all $\Delta$. This qualitative difference between
glycerol and water in the origin of the Nusselt number modification also means that the experimental 
finding $F_{\lambda} \approx 1$  for water at $T_m=40K$ and $Ra$ in the range of $10^8-10^{10}$, see 
ref.\cite{ahl06}, is merely accidental and not a general feature of the RB flow under NOB conditions.

Both NOB responses, the shift of the center temperature $T_c$ and thus $\Delta_b \neq \Delta_t$ as well as 
the shift of the BL thicknesses $\lambda^{sl}_{b,t}$, are determined by the full nonlinear dynamics, in glycerol as 
well as in water. The $T_c$-shift in glycerol is even larger ($\approx 6.5$K) than in water ($\approx 1$K). The same
is expected for the $\lambda^{sl}_{b,t}$-shifts. But the differences in the temperature drops $\Delta_{b,t}$ 
enter via $F_{\Delta}$; here they are weighed with the explicit temperature dependence of the material 
parameter $\kappa(T)$.
Since the thermal diffusivity changes only minutely in glycerol, $\kappa_{b,t} / \kappa_m -1 \approx \pm 0.01$, the
factor $F_{\Delta}$ stays near $F_{\Delta} \approx 1$ despite the large $T_c$ response, cf. 
Figs.~\ref{glyc-f-vs-del},\ref{glyc-nu-vs-del}. This does not happen in $F_{\lambda}$; here the full changes of 
$\lambda^{sl}_{b,t}$ enter. Because of the very strong and in particular nonlinear temperature dependence of $\nu$ 
the thicknesses of the BLs change significantly and also quite differently in magnitude at the bottom and the 
top BLs, because $\sqrt{\nu_t / \nu_b} \approx 2.4$ due to the strong nonlinear $T$-dependence of $\nu(T)$. 
Therefore the sum $\lambda^{sl}_b + \lambda^{sl}_t$ no longer is equal to $2 \lambda^{sl}_{OB}$. For water, 
instead, the dominantly linear $\lambda^{sl}_{b,t}$-NOB modifications
are opposite in sign and nearly cancel in the sum of the NOB thicknesses, giving
$\lambda_b^{sl} + \lambda^{sl}_t \approx 2 \lambda^{sl}_{OB}$ or
$F_{\lambda} \approx 1$. Thus in glycerol we have $F_{\Delta} \approx 1$ 
and the Nu changes are dominated by $F_{\lambda}$, while in water 
it is $F_{\lambda} \approx 1$ and the NOB effects in $Nu$ are determined dominantly by $F_{\Delta}$ (which is given 
by the temperature shift alone).

Figure \ref{glyc-nu-vs-del} shows that the dependences of $Nu_{NOB}/Nu_{OB}$ and $F_{\lambda}$
on the Rayleigh number $Ra$ are non-monotonous. We consider this as due to the nontrivial evolution of various
coherent flow patterns with increasing $Ra$.
In particular, as shown in figure \ref{glyc-f-vs-del}, for $Ra=10^4$ 
the function $F_{\lambda}(\Delta)$ 
shows a qualitatively opposite behavior to that for water, namely $F_{\lambda}$ increases with increasing $\Delta$
and reaches as large a value as $1.017$ at $\Delta=50K$. A consequence of our finding is that in general
$Nu_{NOB}/Nu_{OB}=F_\lambda \cdot  F_\Delta$ {\em cannot} be calculated within the extended BL theory introduced 
in ref.\cite{ahl06}, even if a large-scale wind has formed: Within BL theory only the factor $F_\Delta$ can be
calculated but not the factor $F_\lambda$, for which in general one cannot assume $F_\lambda\approx 1$.

\begin{figure}
\begin{center}
\epsfig{file=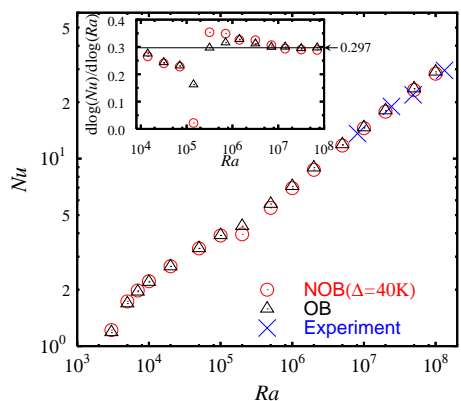,width=5.5cm,angle=-90}
\end{center}
\caption{(color online) 
The Nusselt number $Nu$ for glycerol versus $Ra$ under OB (dashed line) 
and NOB (solid line) 
conditions. In 
both cases $T_m = 40^o$C and $\Delta = 40$K. OB is provided by keeping the material parameters artificially 
constant with $T$. 
We have also included the available data from ref.\cite{zha97}. 
Logarithmic slope ${\rm d}\log(Nu)/{\rm d}\log(Ra)$ is plotted in the
inset and the line corresponding to the exponent $0.297$ measured in
ref.\cite{zha97} is also shown.
} 
\label{glyc-nu-vs-ra}
\end{figure}

We finally present our results for the $Nu$ number itself as a function of $Ra$, see Fig. \ref{glyc-nu-vs-ra}, both for the OB and the NOB case.
  The inset shows the local scaling 
exponents. When applying the unifying theory of refs.\ \cite{gro00},
it is 0.306 at $Ra=10^8$ and $Pr=2500$, consistent with our numerical
findings. This local slope practically does not change in the NOB case.

\section{Summary and conclusions} \label{sec5}
In summary, for glycerol both the center temperature $T_c$ and the Nusselt number $Nu$ 
of the 2D numerical simulations are in good agreement with the available
experimental data of ref.\ \cite{zha97}.  
The experimental finding by Ahlers {\it et al.} \cite{ahl06}
 of a "thermal-BL-thickness sum rule" for 
water, $F_{\lambda} \approx 1$ or $\lambda_b^{sl} + \lambda_t^{sl} \approx 2 \lambda_{OB}^{sl}$, is shown 
to be incidental and seems due to the specific temperature dependence of the material parameters of water 
at $40^{\rm o}{\rm C}$. Apparently this cannot be generalized to other fluids (or other mean temperatures), as 
our analysis of RB convection in glycerol has shown. While for water the Nusselt number modification
$Nu_{NOB}/Nu_{OB}$ is due to the modified temperature drops over the BLs, represented by $F_{\Delta}$, as shown in 
refs.\cite{ahl06,sug07}, for glycerol it is governed by the variation of the BL thicknesses, namely by 
$F_{\lambda}$. This can be attributed to the strong and nonlinear temperature dependence of $\nu (T)$.

\begin{acknowledgments}
\noindent
{\it Acknowledgment:} We thank Guenter Ahlers and Francisco Fontenele Araujo
for many fruitfull discussions over the last years.
The work in Twente is part of the research program of FOM, which is financially supported by NWO and 
SGn acknowledges support by FOM.
\end{acknowledgments}


\end{document}